\documentclass[letterpaper, 10 pt, conference, onecolumn]{ieeeconf} 

\IEEEoverridecommandlockouts                             
\usepackage{graphics} 
\usepackage{epsfig} 
\usepackage{mathptmx} 
\usepackage{times}
\usepackage{amsmath} 
\usepackage{amssymb}  
\usepackage{xcolor}
\usepackage{algpseudocode}
\usepackage{algorithm}
\usepackage{nicefrac, xfrac}
\usepackage{graphicx}
\usepackage{dblfloatfix}
\usepackage{caption}
\usepackage{subcaption}
\usepackage{svg}
\usepackage{array}
\usepackage{hyperref}
\usepackage{todonotes}
\usepackage{caption}

\title{\LARGE \bf
On the Kite-Platform Interactions in Offshore Airborne\\
Wind Energy Systems: Frequency Analysis and Control Approach}

\author{Sofia Trombini, Edoardo Pasta, Lorenzo Fagiano
\thanks{This study was carried out within the NEST - Network 4 Energy Sustainable Transition (D.D. 1243 02/08/2022, PE00000021) and received funding under the National Recovery and Resilience Plan (NRRP), Mission 4 Component 2 Investment 1.3, funded from the European Union - NextGenerationEU. This manuscript reflects only the authors’ views and opinions, neither the European Union nor the European Commission can be considered responsible for them.}
\thanks{Sofia Trombini and Lorenzo Fagiano are with the Dipartimento di Elettronica, Informazione e Bioingegneria (DEIB), Politecnico di Milano, Piazza Leonardo da Vinci 32, 20133 Milano, Italy. e-mail addresses: sofia.trombini@polimi.it; lorenzo.fagiano@polimi.it.Edoardo Pasta is with Marine Offshore Renewable Energy Lab., Dipartimento di Ingegneria Meccanica e Aerospaziale (DIMEAS), Politecnico di Torino, Corso Duca degli Abruzzi
24, 10129 Torino, Italy. e-mail address: edoardo.pasta@polito.it}
\thanks{This article is currently under review for the European Control Conference 2024.}}

\begin{document}

\maketitle
\thispagestyle{empty}
\pagestyle{empty}

\begin{abstract}

This study investigates deep offshore, pumping Airborne Wind Energy systems, focusing on the kite-platform interaction. The considered system includes a 360 $m^2$ soft-wing kite, connected by a tether to a winch installed on a 10-meter-deep spar with four mooring lines.Wind power is converted into electricity with a feedback controlled periodic trajectory of the kite and corresponding reeling motion of the tether.An analysis of the mutual influence between the platform and the kite dynamics, with different wave regimes, reveals a rather small sensitivity of the flight pattern to the platform oscillations; on the other hand, the frequency of tether force oscillations can be close to the platform resonance peaks, resulting in possible increased fatigue loads and damage of the floating and submerged components. A control design procedure is then proposed to avoid this problem, acting on the kite path planner. Simulation results confirm the effectiveness of the approach.\\

\end{abstract}

\section{INTRODUCTION}
Airborne Wind Energy Systems (AWES) convert high-altitude wind energy using a tethered aircraft, or kite \cite{VERMILLION,fagiano2022}. 
AWES can reach altitudes higher than 300 m above ground, where winds are strong with high probability, yielding large capacity factors \cite{BECHTLE20191103}. Moreover, they can be manufactured, transported and installed at low cost, thanks to the absence of large monolithic components, making them appealing for remote locations. These features make AWES a strong candidate technology to complement traditional wind energy and solar PV and increase the overall penetration of renewables in our energy mix, towards the goals of net-zero emissions set by most countries \cite{BVG2022}. The main drawback of AWE technologies is the rather high operational complexity, mostly residing in the automation and control system \cite{fagiano2022}.
In the past 20 years, AWES development has significantly increased. Today, pre-series production of onshore and inland systems in the 100-kW range, grid connected in remote locations, has started, and there is a well-established community of companies and academic institutions that are conducting extensive research on the various involved aspects \cite{Schmehl2018,fagiano2022}. 
Most activities pertain to inland systems, which are more accessible and cheaper for research and development than offshore ones. However, the latter hold the promise of a huge potential impact, in particular for deep offshore locations, where the key advantages of AWES can pave the way to economic viability and environmental sustainability at large scale. In fact, the expected mass of the floating platform for AWES is a rather small fraction than the one required by offshore horizontal-axis wind turbines, thanks to the center of gravity and applied force being close to the sea surface, thus abating the transportation and installation costs. Moreover, the offshore wind resources are abundant already below 300$\,$m, so that, thanks to the possibility to harvest energy at different altitude layers (thus limiting the wake interactions), AWES farms can be arranged compactly, reaching a rather high unit density per occupied surface area \cite{5350676}.
Notwithstanding its promising features, there are currently very few contributions in the literature on deep offshore AWES. In \cite{Cherubini-Fontana-short} and \cite{Cherubini-Fontana-long}, this concept is evaluated using a simplified model. The results indicate technical feasibility, yet considerable platform displacements are observed, while the kite-floater interactions are not treated in detail. Regarding real-world installations, the company Makani Power \cite{Makani} attempted a medium-scale (500~kW) offshore system deployment in Norway in 2019. Unfortunately, the company  operations ended after a few months.\\
This paper contributes to advance the knowledge on offshore AWES, focusing on the kite-platform interaction. A pumping AWE system is considered, which converts wind power
into electricity with a feedback-controlled periodic trajectory
of the kite and corresponding reeling motion of the tether. In contrast to \cite{Cherubini-Fontana-long}, the kite model is not mass-less and the tether is a nonlinear spring with elastic constant depending on its length, whereas in \cite{Cherubini-Fontana-long} it was assumed to be a rigid rod.  The first contribution is an analysis of the mutual influence between the platform and the kite dynamics, with different wave regimes, using a 6-degrees-of-freedom (d.o.f.) model of the platform coupled with an established model of the AWE system. We find that the frequency of tether force oscillations
can be close to the platform resonance peaks, resulting in possible increased fatigue loads and damage of the floating and submerged components. The second contribution is to propose a control design procedure to avoid this problem, acting on the kite path planner.
Simulation results confirm the effectiveness of the approach.

\section{SYSTEM DESCRIPTION AND MODEL}\label{ch:sys_description}

\begin{figure}[!htb]
    \centering
    \includegraphics[width=13cm]{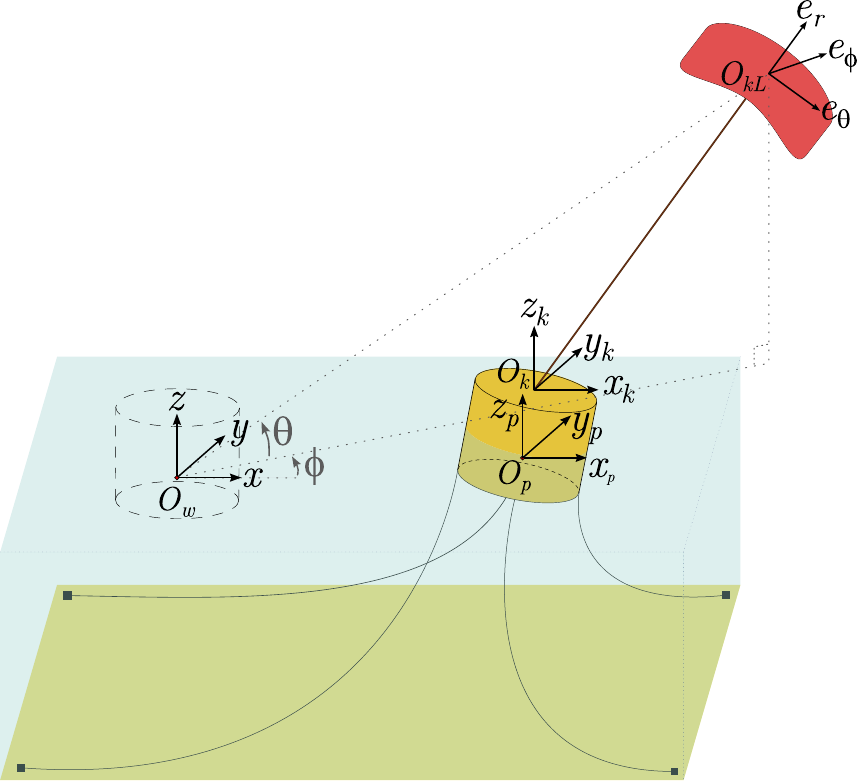}
    \caption{Conceptual layout of the system, with the four considered reference frames.}
    \label{fig:system_1}
\end{figure}

We consider a pumping Airborne Wind Energy system installed on a moored spar-buoy, see Fig. \ref{fig:system_1} for a conceptual layout. We consider a soft kite with one tether, similar to those employed by the company Skysails Power \cite{ERHARD201513,fagiano2022}, and a simple geometry of the spar-buoy (i.e., a cylinder) partially filled with heavy sand to act as a ballast. We further assume the presence of four symmetric catenary moorings to anchor the platform to the seabed.\\ 
four reference frames: 
\begin{itemize}
    \item Fixed, inertial frame ($x, y, z$), with origin $O_{W}$ in the platform centre of gravity at rest, i.e. in static equilibrium when no forces other than gravity and buoyancy are acting on it. The $z$-axis points upwards, perpendicular to the sea surface. 
    \item Platform frame ($x_{P}, y_{P}, z_{P}$) with origin $O_{P}$ in the platform centre of gravity and axes coinciding at rest with the inertial one. 
    \item AWES platform frame ($x_{K}, y_{K}, z_{K}$) with origin $O_{K}$ located at the tether exit point from the platform. At rest, the axes' are parallel to those of the inertial frame.
    \item AWES local reference system ($e_{\theta}, e_{\phi}, e_{r}$) with origin $O_{KL}$ in the wing. The unit vectors ($e_{\theta}, e_{\phi}, e_{r}$) are defined in the fixed reference system ($x, y, z$) as 
    \begin{equation}
    \begin{array}{c}
    \begin{bmatrix}
        e_{\theta} & e_{\phi} & e_{r}
    \end{bmatrix}
    =
    \begin{bmatrix}
        \sin(\theta)\,\cos(\phi) & -\sin(\phi) & \cos(\theta)\,\cos(\phi) \\
        \sin(\theta)\,\sin(\phi) & \cos(\phi) & \cos(\theta)\,\sin(\phi) \\
        -\cos(\theta) & 0 & \sin(\theta) 
    \end{bmatrix}
    \end{array}
    \nonumber
    \end{equation}
    where  $\theta$ and $\phi$, named elevation and azimuth, respectively, are the angles describing the wing's position in the inertial frame using polar coordinates (see Fig. \ref{fig:system_1}).
\end{itemize}

\noindent The next subsections provide the mathematical equations for each system's component. 
\\[5pt]

\subsection{Airborne wind energy system model} \label{AWES model}
We adopt the point-mass AWES model of  \cite{5152910}, briefly recalled here for the sake of self-consistency. The kite's position relative to the origin of the inertial reference frame can be expressed using the distance $r$ from $O_W$ to $O_{KL}$ and angles $\theta$ and $\phi$:
\begin{equation}
    \Vec{p}_{W-KL}(t)=\begin{bmatrix}
        r(t)\,\cos(\theta(t))\,\cos(\phi(t)) \\ r(t)\,\cos(\theta(t))\,\sin(\phi(t)) \\ r(t)\,\sin(\theta(t))
    \end{bmatrix}
    \nonumber
\end{equation}
where $t\in\mathbb{R}$ is the continuous-time variable.
Newton's law of motion in the AWES local reference system yields the equations:
\begin{equation}\label{eq:kite_equations}
\begin{aligned}
    &\Ddot{\theta}(t)=\frac{F_{\theta}(t)}{m(t)\,r(t)} \\
    &\Ddot{\phi}(t)=\frac{F_{\phi}(t)}{m(t)\,r(t)\,\sin(\theta(t))} \\
    &\Ddot{r}(t)=\frac{F_{r}(t)}{m(t)}
\end{aligned}
\end{equation}
where $m(t)$ is the kite's mass $m_K$ augmented by half of the tether mass as (see \cite{5152910}):
\[
m(t)=m_K+\dfrac{\rho_t\pi d_t^2 L(t)}{8}
\]
with $\rho_t$, $d_t$, and $L(t)$ being, respectively, the tether density, its diameter, and its nominal (i.e., with zero pulling force) length measured from the exit point on the floating platform to the kite. 
$F_{\theta}(t)$, $F_{\phi}(t)$ and $F_{r}(t)$ in \eqref{eq:kite_equations} are the resultant forces for each axis of the local reference frame. In particular (variable $t$ is omitted for simplicity):
\begin{equation}
\begin{aligned}
    & F_{\theta} = F_{\theta}^{grav} + F_{\theta}^{app} + F_{\theta}^{aer}\\
    & F_{\phi} = F_{\phi}^{grav} + F_{\phi}^{app} + F_{\phi}^{aer}\\
    & F_{r} = F_{r}^{grav} + F_{r}^{app} + F_{r}^{aer} - F^{c,trc} \\
\end{aligned}
\nonumber
\end{equation}
where $\Vec{F}^{grav}(t)$ is the gravity force, $\Vec{F}^{app}(t)$ the apparent force, $\Vec{F}^{aer}(t)$ the kite aerodynamic force, and $\Vec{F}^{c,trc}(t)$ the tether traction force. The aerodynamic forces are nonlinear functions of the wind speed vector at the kite position, denoted by $\Vec{W}(t)$, and depend on the kite effective area $A$ and lift and drag coefficients, $C_L,\,C_D$ accounting also for the tether drag, which in turn depends on its diameter and length, see e.g. \cite{5152910} for the details. In this work, the tether is modelled as a nonlinear spring exerting a pulling force equal in absolute value to:
\begin{equation}\label{eq:tether_force_mag}
    F^{c,trc}(t)=\max\left(0,k(L(t))\,\|\Vec{P}_{K-KL}(t)\|_2-L(t)\right)
\end{equation}
where $k(L(t))$ is the spring coefficient and $\Vec{P}_{K-KL}(t)$ is the vector pointing from $O_{K}$ to $O_{KL}$, computed as:  
\begin{equation}\label{eq:vector-kite-relative-to-exit point}
    \Vec{P}_{K-KL}(t)=\Vec{P}_{W-KL}(t) - \Vec{P}_{W-K}(t),
\end{equation}
such that $\|\Vec{P}_{K-KL}(t)\|_2$ represents the distance between the kite and the tether exit point on the platform. In \eqref{eq:vector-kite-relative-to-exit point}, $\Vec{P}_{W-K}(t)$  is the vector pointing from $O_{W}$ to $O_{K}$, computed as:
\begin{equation}\label{eq:vector_sum}
    \Vec{P}_{W-K}(t)=\Vec{P}_{W-O}(t) + \Vec{P}_{O-K}  
\end{equation}
i.e., the vector sum of $\Vec{P}_{W-O}(t)$, pointing from the origin of the inertial plane to the platform's center of gravity, and $\Vec{P}_{O-K}$, pointing from the latter to the exit point of the tether. Vector $\Vec{P}_{O-K}$ is assumed to be constant, since the platform is modeled as a rigid body, while $\Vec{P}_{W-O}(t)$ depends on the position of the platform, as specified later on. About the spring coefficient $k(L(t))$ in \eqref{eq:tether_force_mag}, this is computed as:
\[
    k(L(t))= \frac{\Bar{F}^{c,trc}}{\Bar{\epsilon}\, L(t)}
\]
where $\Bar{F}^{c,trc}$ is the breaking load of the tether, and $\Bar{\epsilon}$ the corresponding elongation (e.g., $\Bar{\epsilon}=0.03$ means that the tether reaches the breaking load when $\|\Vec{P}_{K-KL}(t)\|_2$ equals 1.03 times the nominal tether length, $L(t)$).
Regarding the flight control approach, the selected strategy is the one described in \cite{AWEmodel}. It features a hierarchical structure, where a high-level navigation approach employs two user-defined target points, $P_{-}=(\theta_{-},\phi_{-})$ and $P_{+}=(\theta_{+},\phi_{+})$ in the ($\theta$,$\phi$) plane, to compute a reference course for the kite at each discrete time step $k \in \mathbb{Z}$ of the digital control loop. The course is characterized by the so-called velocity angle $\gamma(t)$, defined as:
\begin{equation}\label{eq:gamma}
\gamma(t)=\arctan{ \left( \frac{\Dot{\phi}(t) \cos{(\theta(t))} }{\Dot{\theta}(t)} \right) }
\nonumber
\end{equation}
The following switching algorithm selects the active target point, where it is assumed that $\phi_{-}<\phi_{+}$:
\begin{algorithm}[H]
\caption{Active target point selection}\label{alg:target point}
\begin{algorithmic}
\If {$\phi(k)<\phi_{-}$}
    \State $(\theta_a(k),\phi_a(k))=P_{+}$
\ElsIf {$\phi(k)>\phi_{+}$}
    \State $(\theta_a(k),\phi_a(k))=P_{-}$
\Else
    \State $(\theta_a(k),\phi_a(k))=(\theta_a(k-1),\phi_a(k-1))$
\EndIf
\end{algorithmic}
\end{algorithm}
Then, the desired velocity angle is computed as:
\begin{equation}
    \gamma_{ref}(k)=\arctan\left(\frac{(\phi_a(k)-\phi(k))\,\cos(\theta(k))}{\theta_a(k)-\theta(k)}\right)
\nonumber
\end{equation}
This reference corresponds to a kite velocity vector that points towards the active target point in the ($\theta$,$\phi$) plane. At the lower level, a proportional controller computes the kite's steering input $\delta(k)$ in order to track $\gamma_{ref}(k)$. 
More details on this kite control strategy can be found in \cite{AWEmodel}.

\subsection{Platform model}\label{ss:platform}

The  motion of objects surrounded by a fluid, like the floating platform considered in this work, is commonly described using the Navier-Stokes' equations \cite{drazin_riley_2006}. 
However, their numerical solution is computationally expensive, and, for this reason, not suitable for parametric analysis of the dynamics of complex interconnected systems or for control-oriented modelling. For this reason, we consider a 6-d.o.f. model based on the linear potential flow theory \cite{cummins}, which introduces some simplifications but still provides reliable results. In this framework, the fluid surrounding the body is incompressible and inviscid, and the flow is irrotational \cite{Faedo}. To resolve the dynamics and fulfil boundary requirements on the platform, boundary element techniques (BEMs) are used in conjunction with the linear potential flow theory. In this study, we used the NEMOH programme as a numerical solver to evaluate the hydrodynamic parameters \cite{Nemoh}. \\
The platform and its mooring are modeled as a mass-spring-damper system with the addition of specific hydrodynamic forces and moments; the resulting six equations of motion are: 
\begin{equation}\label{eq:platform_motion}
    M\,\Ddot{\nu}(t) =F^{h}(t) + F^{r}(t) + F^{exc}(t) + F^{m}(t) + F^{t}(t) 
\end{equation}
where $\nu(t)=[x_P(t) \; y_P(t) \; z_P(t) \; \omega_{x_P}(t) \; \omega_{y_P}(t) \; \omega_{z_P}(t)]^{T}$ are the displacements (i.e., surge, sway and heave) and rotations with respect to the platform centre of gravity, $M\in\mathbb{R}^{6\times 6}$ is the mass-inertia matrix,  vector $F^{h}\in\mathbb{R}^{6}$ contains the three-dimensional hydrostatic restoring force and moment, $F^{r}$ the radiation force and moment, $F^{exc}$ the wave excitation force and moment, $F^{m}$ the mooring force and moment and $F^{t}$ the tether traction force and moment applied on the platform.\\
The hydrostatic restoring effect $F^{h}$ accounts for the static pressure and gravity force, 
 and it is expressed as:
\begin{equation}
    F^{h}(t) = - K_{h}\,\nu(t)
\nonumber
\end{equation}
where $K_{h}$ is the restoring coefficient matrix. The radiation force and moment $F^{r}$ are those exerted by the fluid on the platform when no incident waves are present. Its effect is described as:
\begin{equation}
    F^{r}(t) = -M_{\infty}\,\Ddot{\nu}(t) -\int_{-\infty}^{t}h_{ra}(t-\tau)\Dot{\nu}(\tau)\,d\tau
\nonumber
\end{equation}
where $M_{\infty}$ is a 6x6 matrix accounting for the added mass (see \cite{added-mass}), and $h_{ra}(t)$ is the  impulse response of the radiation dynamics, which accounts for the memory effect due to the fluid action. The wave excitation $F^{exc}$ accounts for the impact of waves on the platform. Following the linear potential flow theory assumptions, this force results from the superimposition of the Froude-Krylov force and diffraction: the first is the consequence of wave pressure on a so-called ``ghost'' body that doesn't influence the wave field but responds to it, while the second takes into account body interference on the wave field \cite{Faedo,Giorgi}.  Finally, since the mooring lines are modeled as mass-spring-dampers, the mooring force $F^{m}$ is:
\begin{equation}
    F^{m}(t) = -M_{m}\,\Ddot{\nu}(t) - B_{m}\,\Dot{\nu}(t) - K_{m}\,\nu(t)
\nonumber
\end{equation}
where $M_{m}$ is the mooring inertia matrix, $B_{m}$ is the mooring damping matrix and $K_{m}$ is the mooring stiffness matrix. These parameters are obtained through a system identification process using data from Orcaflex software. Finally, the tether traction force and moments $F^{t}$ are computed as:
\begin{equation}\label{eq:tether_force_platform}
F^{t}=\left[
\begin{array}{c}
\Vec{F}^{c,trc}\\
\Vec{F}^{c,trc}\times\Vec{P}_{O-K}
\end{array}
\right]
\end{equation}
where $\times$ denotes the cross-product, and 
\[
\Vec{F}^{c,trc}=F^{c,trc}\dfrac{\Vec{P}_{K-KL}}{\|\Vec{P}_{K-KL}\|_2}\\
\]
with $F^{c,trc}$ computed as in \eqref{eq:tether_force_mag}.

\subsection{Overall model equation}
The AWES' and platform's dynamical models are coupled via the tether force \eqref{eq:tether_force_platform}, which acts on the platform motion \eqref{eq:platform_motion} and depends on the platform position via the equation:
\[
\Vec{P}_{W-O}(t)=\left[
\begin{array}{c}
x_P(t)\\y_P(t)\\z_P(t)
\end{array}\right],
\]
see \eqref{eq:tether_force_mag}-\eqref{eq:vector_sum}. Thus, we can express the overall system equations as
\begin{equation}
    \dot{x(t)}=f(x(t),F^{exc}(t),\Vec{W}(t))  
\nonumber
\end{equation}
where $ x(t) = [\theta(t) \; \phi(t) \; r(t) \; \Dot{\theta}(t) \; \Dot{\phi}(t) \; \Dot{r}(t) \; x_P(t) \; y_P(t) \; z_P(t) \; \omega_{x_P}(t) \; \omega_{y_P}(t) \; \omega_{z_P}(t)\; \Dot{x_P}(t) \; \Dot{y_P}(t) \; \Dot{z_P}(t) \; \Dot{\omega_{x_P}}(t)\; \Dot{\omega_{y_P}}(t) \; \Dot{\omega_{z_P}}(t)]^T$ and $F^{exc}(t),\,\Vec{W}(t)$
are exogenous inputs.

\section{ANALYSIS OF KITE-PLATFORM INTERACTIONS}\label{ch:analysis}

Using the described model, we carried out an analysis of how the platform motion affects the kite's behavior, and vice-versa. The results presented here have been obtained with the model parameters reported in Table \ref{tab:model}, corresponding to a medium-size AWES with average cycle power of 500~kW, and considering the traction phase of the pumping cycle, when the  flight controller described in Section \ref{ch:sys_description} is active. We estimated the AWES parameters by scaling up those of the Skysails Power SKS PN-14 reported in \cite{fagiano2022}, considering that the aerodynamic forces grow linearly with the kite surface, assuming a constant kite mass per unit area, and scaling accordingly the tether diameter to have a breaking load equal to 4 time (safety coefficient) the maximum expected traction force values. Regarding the buoy mass, it corresponds to a spar with diameter and height of 10~m. When no force is acting on it, the height of the platform out of the water is approximately 1~m. The ballast height is 3.6~m. The buoy centre of gravity lies on its symmetry axis, about 7~m below the deck, and its mass is about 760 tons. The numerical values of parameters $M,\,K_h,\,M_\infty,\,M_m,\, B_m,$ and $K_m$ and the impulse response $h_{ra}(t)$ introduced in Section \ref{ss:platform} are omitted here due to space limitations, however we made available a Matlab file containing them (where $h_{ra}(t)$ is given via a state-space realization), see \cite{parametri}.\\
Regarding the exogenous inputs, the wave excitation $F^{exc}(t)$ is described using the JONSWAP spectrum \cite{JONSWAP}, characterised by three main parameters: the significant wave height $H_s$, the peak wave period $T_e$ and the peak-shape parameter $\gamma_j$. We considered two sets of JONSWAP spectrum parameters, A and B, where B corresponds to higher waves (see Table \ref{tab:waves}). The free surface elevation and the excitation force are computed using a random amplitude scheme, as explained in Merigaud \cite{Merigaud}. For the wind speed $W(t)$, we considered instead a uniform wind field directed along the inertial $x$-axis, with magnitude equal to 8.5 m/s. This is a realistic value for the relative wind speed (i.e., absolute wind speed minus the tether reel-out speed) experienced by the kite during the traction phase of the pumping cycle above rated power, when the maximum tether force values are reached. We did not include a wind turbulence model in order to isolate the dynamical effects related to the wave excitation.\\
We first focus on the tether force behavior and compare the outcome with that of an onshore scenario.
\begin{table}[h]
\centering
\caption{Model parameters employed in the analysis}
\label{tab:model}
\begin{tabular}{|m{0.5\columnwidth}|m{0.1\columnwidth}|m{0.25\columnwidth}|}
\hline
Kite effective area & $A$ & 360~m$^2$                \\ \hline
Kite mass  & $m_K$ & 90~kg             \\ \hline
Tether diameter  & $d_t$ & 0.035~m             \\ \hline
Tether density  & $\rho_t$ & 980~kg m$^{-3}$             \\ \hline
Tether breaking load  & $\Bar{F}^{c,trc}$ & 490~kN          \\ \hline
Tether breaking elongation  & $\Bar{\epsilon}$ & 0.03             \\ \hline
Tether exit point relative to the 
platform's c.o.g., in frame ($x_{P}, y_{P}, z_{P}$), when platform is at rest & $\Vec{P}_{O-K}(t)$ & $[0,0,7.8475]^T $            \\ \hline
Target points & $P_{-}$,$P_{+}$ & (0.6,-0.4),(0.6,0.4)  \\ \hline
\end{tabular}
\end{table}

\begin{table}[h]
\centering
\caption{JONSWAP parameters for the two waves}
\label{tab:waves}
\begin{tabular}{|c|c|c|c|}
\hline
\textbf{}       & \textbf{$H_{s}$ ($m$)} & \textbf{$T_e$ ($s$)} & \textbf{$\gamma_j$} \\ \hline
\textbf{wave A} & 0.5 & 3.7 & 3.1                \\ \hline
\textbf{wave B} & 2 & 7.5 & 3.1                \\ \hline
\end{tabular}
\end{table}

\begin{figure}[h]
    \begin{subfigure}{\textwidth}
    \centering
        \includegraphics[width=0.65\textwidth]{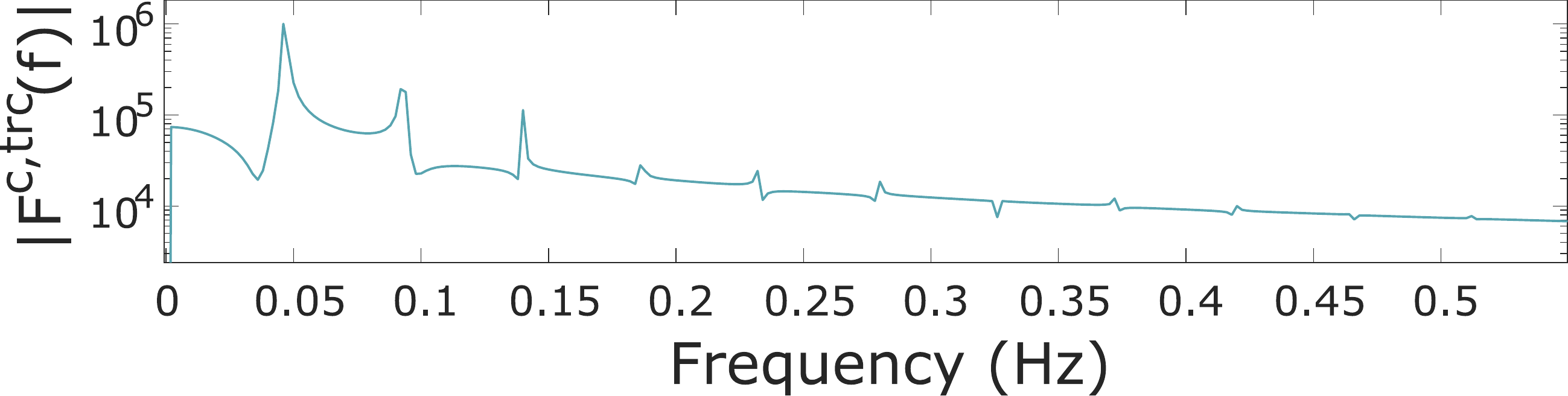}
        \caption{}
        \label{subfig:onshore_spectrum}
    \end{subfigure}
    \\
    \\
    \begin{subfigure}{\textwidth}
    \centering
        \includegraphics[width=0.65\textwidth]{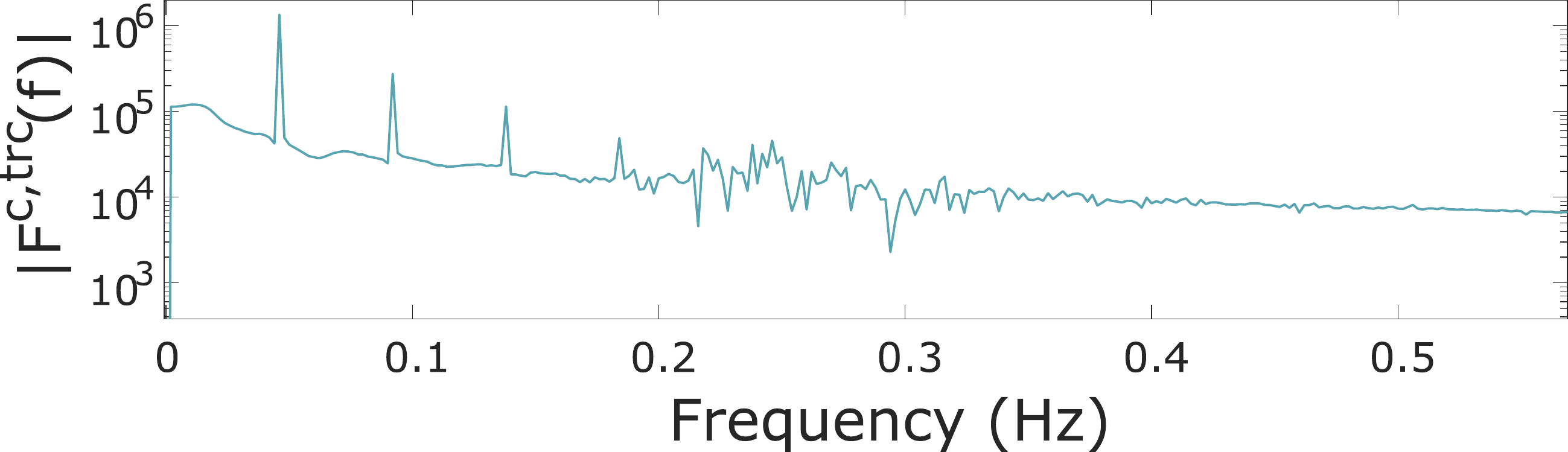}
        \caption{}
        \label{subfig:waveA_spectrum}
        \end{subfigure}
    \\
    \\
    \begin{subfigure}{\textwidth}
    \centering
        \includegraphics[width=0.65\textwidth]{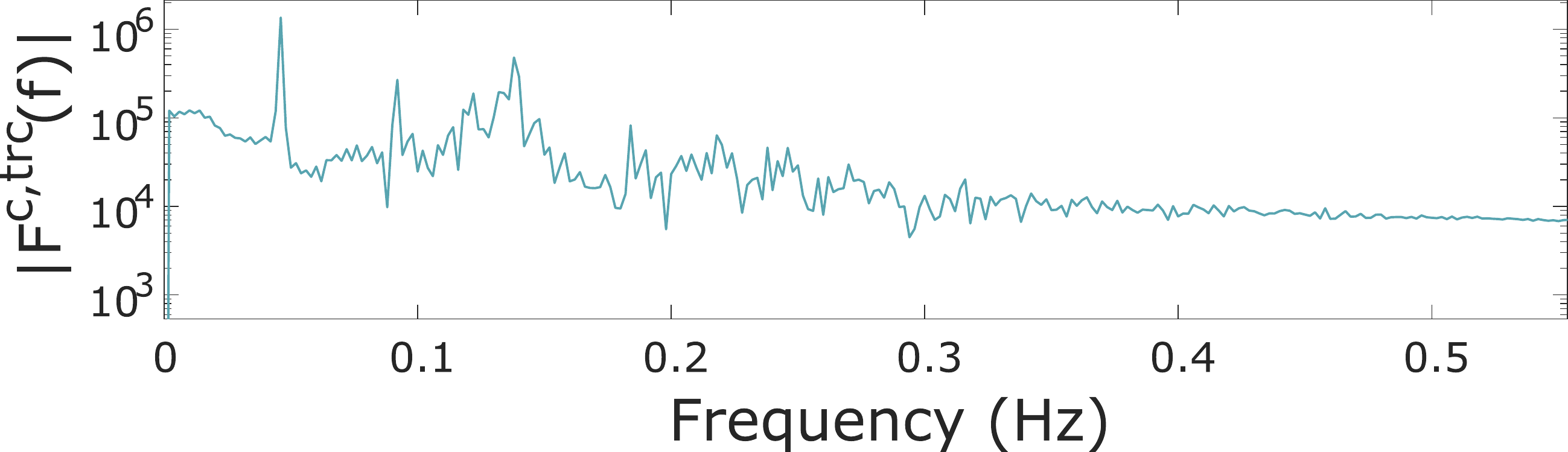}
        \caption{}
        \label{subfig:waveB_spectrum}
    \end{subfigure}
    \caption{Tether force spectrum with tether length $L=900\,$m. (a) Onshore. (b) Offshore - Wave A. (c) Offshore - Wave B. }
    \label{subfig:total_spectrum}
\end{figure}

\noindent Figs. \ref{subfig:onshore_spectrum}-\ref{subfig:waveB_spectrum}  show the tether force spectra when $L=900\,$m. In the onshore case (Fig. \ref{subfig:onshore_spectrum}), the spectrum contains only the frequencies pertaining to the kite's motion, while looking at wave A and wave B scenarios in Figs. \ref{subfig:waveB_spectrum} and \ref{subfig:waveB_spectrum}, we note the presence of additional components due to the waves, respectively around 0.22-0.3~Hz for case A and above 0.1-0.15~Hz for case B, the latter being significantly larger. We further studied the effects of waves on the tether force by evaluating its mean value, $F_{mean}$, the mean of its peaks, $Peaks_{mean}$, the force amplitude, $\Delta F$, and the standard deviation of the peaks, $std_{peaks}$ over 100 periodic patterns. Table \ref{tab:wave on tether} presents the results, reported in kN, showing that the average force is practically unaffected, while its variability and peaks change in a very limited way in case A and in a much stronger one in case B. 
\\[5pt]

\begin{figure}[h]
    \centering
    \includegraphics[width=0.65\textwidth]{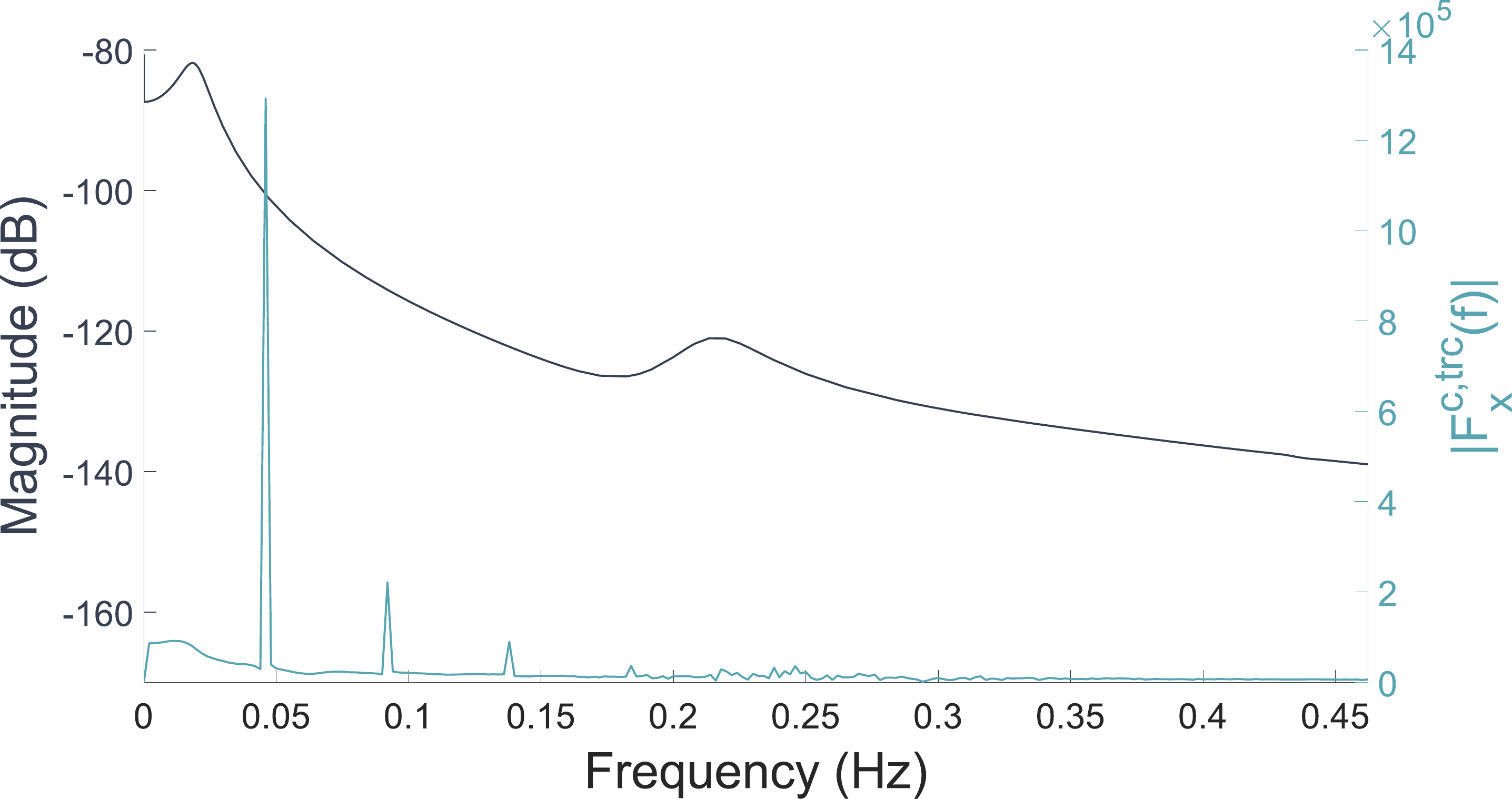}
    \caption{Bode diagram of the frequency response of $x_P$ w.r.t the traction force $x$-component, and spectrum of the latter for wave A and tether length $L=900\,$m.}
    \label{fig:bode_in1_out1}
\end{figure}

\begin{figure}[H]
    \centering
    \includegraphics[width=0.65\textwidth]{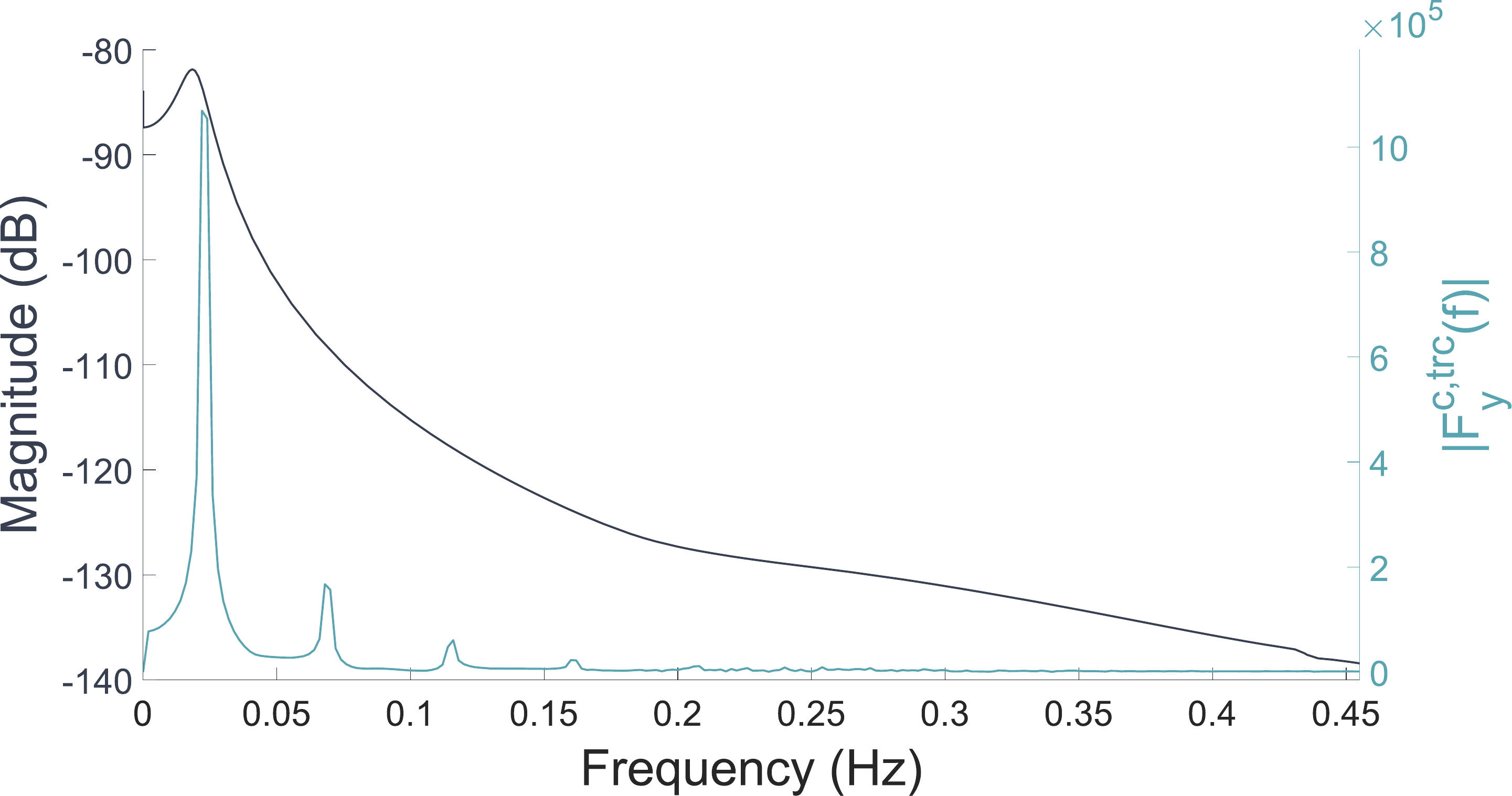}
    \caption{Bode diagram of the frequency response of $y_P$ w.r.t the traction force $y$-component, and spectrum of the latter for wave A and tether length $L=900\,$m.}
    \label{fig:bode_in2_out2}
\end{figure}

\begin{figure}[H]
    \centering
    \includegraphics[width=0.65\textwidth]{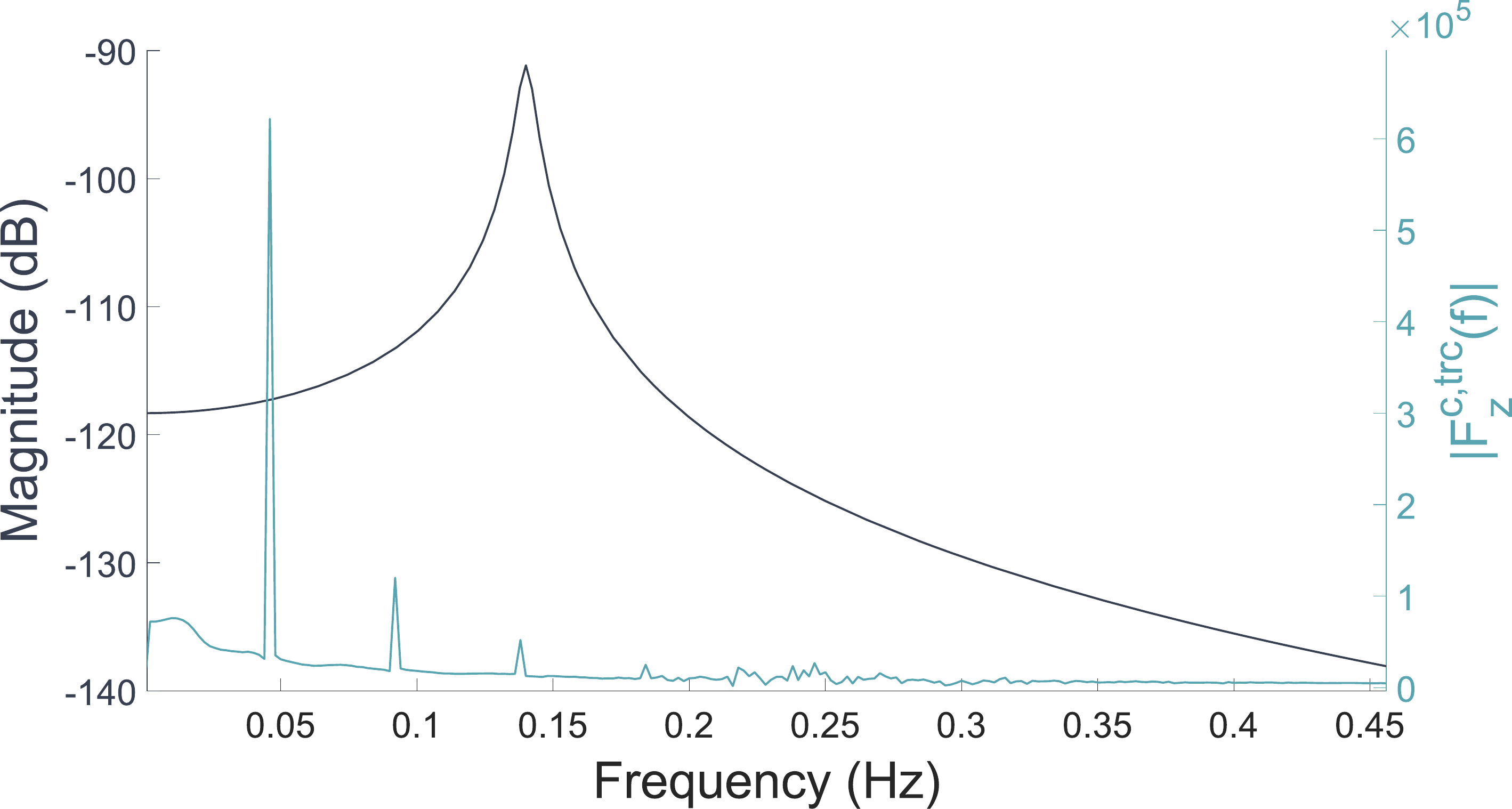}
    \caption{Bode diagram of the frequency response of $z_P$ w.r.t the traction force $z$-component, and spectrum of the latter for wave A and tether length $L=900\,$m.}
    \label{fig:bode_in3_out3}
\end{figure}

\noindent Next, we carried out a frequency response analysis on the platform, to assess whether the main spectral components of tether force are close to the platform's resonance peaks. Figures \ref{fig:bode_in1_out1}, \ref{fig:bode_in2_out2}, and \ref{fig:bode_in3_out3} show the Bode diagrams of the frequency response function of $x_P$, $y_P$, and $z_P$ with respect to the $x,\,y$ and $z$ components of the tether force vector, respectively, with wave A and tether length equal to 900 m. The highest resonance peaks are at $r_1=r_2=0.0185\;$Hz for $x_P$ and $y_P$ and $r_3=0.14\;$Hz for $z_P$. The peak $r_3$ results to be much higher than the main frequency of oscillation of the vertical tether force, presented in Fig. \ref{fig:bode_in3_out3}, hence, we expect a limited increase and no resonance effects for the platform's heave motion. On the other hand, the resonances $r_1$ and $r_2$ are rather close to the principal components of the tether force in the $x_P,\,y_P$ directions, see Figs. \ref{fig:bode_in1_out1}-\ref{fig:bode_in2_out2}. In particular, the $y$ component of the tether force, which has the same frequency as that of the kite's eight-path, results to be very close to the first platform resonance in the sway motion. 
\noindent Moreover, under the employed flight control strategy, whose switching conditions are based on the elevation and azimuth angles only (see Algorithm \ref{alg:target point}), these results are affected by the tether length $L$. In fact, with higher $L$ values, the length of the flown paths increases (because the angular span is the same), while the kite speed decreases, due to the higher tether drag. As a consequence, the frequency of oscillation of the tether force decreases. The overlap between the tether force spectrum and the platform resonance peaks thus depends also on $L$; to study this aspect for the sway motion (i.e., along the $y$ direction) of the platform we introduce the quantity $\eta$, defined as:
\begin{equation}\label{eq:eta}
    \eta \doteq \frac{y_{P_{peak}}}{F^t_{y_{peak}}}
\end{equation}
where $y_{P_{peak}}$ is the peak of the sway oscillation and $F^t_{y_{peak}}$ that of the  tether force component along the $y$ axis. The values of $\eta$ with different tether lengths are shown in  Table \ref{tab:eta}, considering the case of wave A parameters: as expected by intuition,  a significant increase of $\eta$ is observed when the kite trajectory's frequency approaches the resonance $r_2$ In particular, notwithstanding a decrease of the peak force of about 33\% between $L=600\,$m and 1100~m, an increase of 230\% of the platform's sway peak takes place, with a lateral movement spanning the interval $\pm4.6\,$m.

\begin{table*}[t]
\centering
\caption{Comparison of the tether force values obtained in the three considered scenarios. All values in the table are in kN.}
\label{tab:wave on tether}
\begin{tabular}{|c||cccc||cccc||cccc|}
\hline
\textbf{Tether}       & \multicolumn{4}{c||}{\textbf{Onshore}}                                                                                                                                     & \multicolumn{4}{c||}{\textbf{Offshore - wave A }}                                                                                                                           & \multicolumn{4}{c|}{\textbf{Offshore - wave B}}                                                                                                                           \\ \cline{2-13}
\multicolumn{1}{|l||}{\textbf{length} $L$ (m) } & \multicolumn{1}{l|}{\textbf{$F_{mean}$}} & \multicolumn{1}{l|}{\textbf{$Peaks_{mean}$}} & \multicolumn{1}{l|}{\textbf{$\Delta F$}} & \multicolumn{1}{l||}{\textbf{$std_{peaks}$}} & \multicolumn{1}{l|}{\textbf{$F_{mean}$}} & \multicolumn{1}{l|}{\textbf{$Peaks_{mean}$}} & \multicolumn{1}{l|}{\textbf{$\Delta F$}} & \multicolumn{1}{l||}{\textbf{$std_{peaks}$}} & \multicolumn{1}{l|}{\textbf{$F_{mean}$}} & \multicolumn{1}{l|}{\textbf{$Peaks_{mean}$}} & \multicolumn{1}{l|}{\textbf{$\Delta F$}} & \multicolumn{1}{l|}{\textbf{$std_{peaks}$}} \\ \hline
600                    & \multicolumn{1}{c|}{252}                 & \multicolumn{1}{c|}{294}                     & \multicolumn{1}{c|}{42}                  & 0.003                                       & \multicolumn{1}{c|}{249}                 & \multicolumn{1}{c|}{294}                     & \multicolumn{1}{c|}{45}                  & 2                                           & \multicolumn{1}{c|}{250}                 & \multicolumn{1}{c|}{311}                     & \multicolumn{1}{c|}{61}                  & 17                                          \\ 
700                    & \multicolumn{1}{c|}{243}                 & \multicolumn{1}{c|}{286}                     & \multicolumn{1}{c|}{43}                  & 0.003                                       & \multicolumn{1}{c|}{240}                 & \multicolumn{1}{c|}{287}                     & \multicolumn{1}{c|}{47}                  & 2.2                                         & \multicolumn{1}{c|}{240}                 & \multicolumn{1}{c|}{303}                     & \multicolumn{1}{c|}{63}                  & 17                                          \\ 
800                    & \multicolumn{1}{c|}{237}                 & \multicolumn{1}{c|}{280}                     & \multicolumn{1}{c|}{43}                  & 0.0009                                      & \multicolumn{1}{c|}{233}                 & \multicolumn{1}{c|}{281}                     & \multicolumn{1}{c|}{48}                  & 2.5                                         & \multicolumn{1}{c|}{233}                 & \multicolumn{1}{c|}{300}                     & \multicolumn{1}{c|}{77}                  & 14.7                                        \\ 
900                    & \multicolumn{1}{c|}{228}                 & \multicolumn{1}{c|}{271}                     & \multicolumn{1}{c|}{43}                  & 0.002                                       & \multicolumn{1}{c|}{223}                 & \multicolumn{1}{c|}{273}                     & \multicolumn{1}{c|}{50}                  & 2.7                                         & \multicolumn{1}{c|}{224}                 & \multicolumn{1}{c|}{292}                     & \multicolumn{1}{c|}{78}                  & 16.4                                        \\ 
1000                   & \multicolumn{1}{c|}{222}                 & \multicolumn{1}{c|}{265}                     & \multicolumn{1}{c|}{43}                  & 0.0005                                      & \multicolumn{1}{c|}{216}                 & \multicolumn{1}{c|}{266}                     & \multicolumn{1}{c|}{50}                  & 2.13                                        & \multicolumn{1}{c|}{217}                 & \multicolumn{1}{c|}{287}                     & \multicolumn{1}{c|}{70}                  & 15.3                                        \\ 
1100                   & \multicolumn{1}{c|}{216}                 & \multicolumn{1}{c|}{257}                     & \multicolumn{1}{c|}{41}                  & 0.009                                       & \multicolumn{1}{c|}{209}                 & \multicolumn{1}{c|}{258}                     & \multicolumn{1}{c|}{49}                  & 1.93                                        & \multicolumn{1}{c|}{210}                 & \multicolumn{1}{c|}{280}                     & \multicolumn{1}{c|}{70}                  & 16.8                                        \\ 
1200                   & \multicolumn{1}{c|}{206}                 & \multicolumn{1}{c|}{249}                     & \multicolumn{1}{c|}{43}                  & 0.1                                         & \multicolumn{1}{c|}{203}                 & \multicolumn{1}{c|}{250}                     & \multicolumn{1}{c|}{47}                  & 1.63                                        & \multicolumn{1}{c|}{203}                 & \multicolumn{1}{c|}{269}                     & \multicolumn{1}{c|}{76}                  & 13.8                                        \\ 
1300                   & \multicolumn{1}{c|}{198}                 & \multicolumn{1}{c|}{242}                     & \multicolumn{1}{c|}{44}                  & 0.05                                        & \multicolumn{1}{c|}{196}                 & \multicolumn{1}{c|}{243}                     & \multicolumn{1}{c|}{47}                  & 2.59                                        & \multicolumn{1}{c|}{196}                 & \multicolumn{1}{c|}{263}                     & \multicolumn{1}{c|}{77}                  & 20                                          \\ \hline
\end{tabular}
\end{table*}

\begin{figure}[h]
    \begin{subfigure}{\textwidth}
    \centering
        \includegraphics[width=0.65\textwidth]{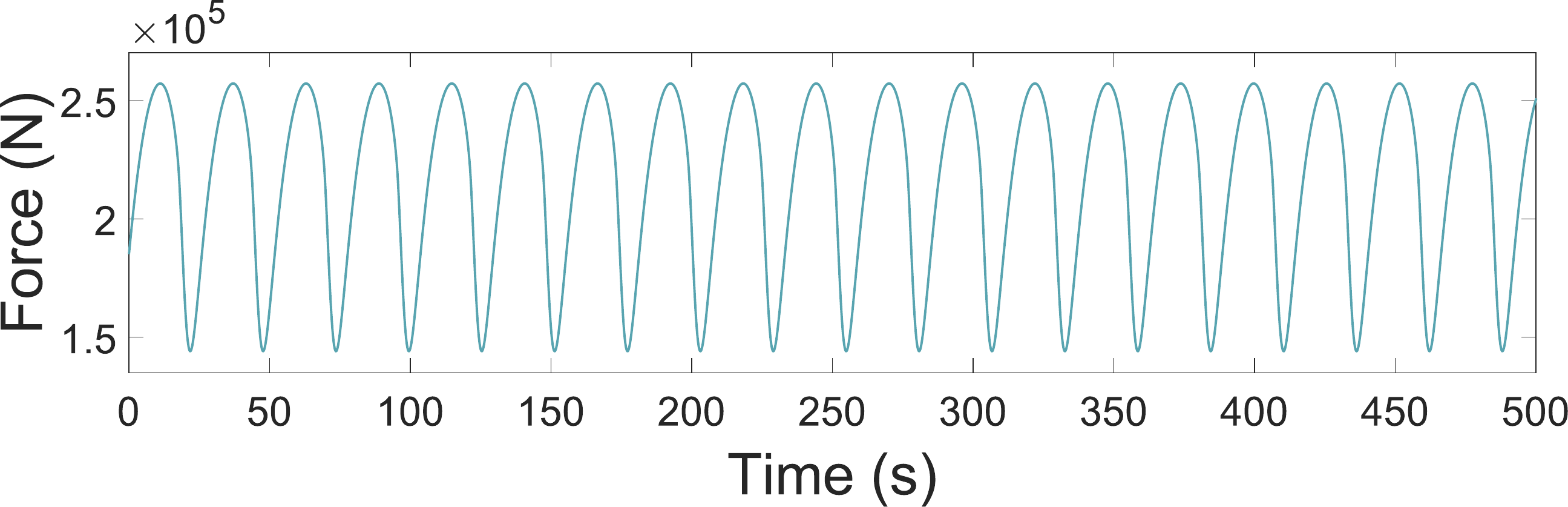}
        \caption{}
        \label{subfig:onshore_forceT}
    \end{subfigure}
    \\
    \begin{subfigure}{\textwidth}
    \centering
        \includegraphics[width=0.65\textwidth]{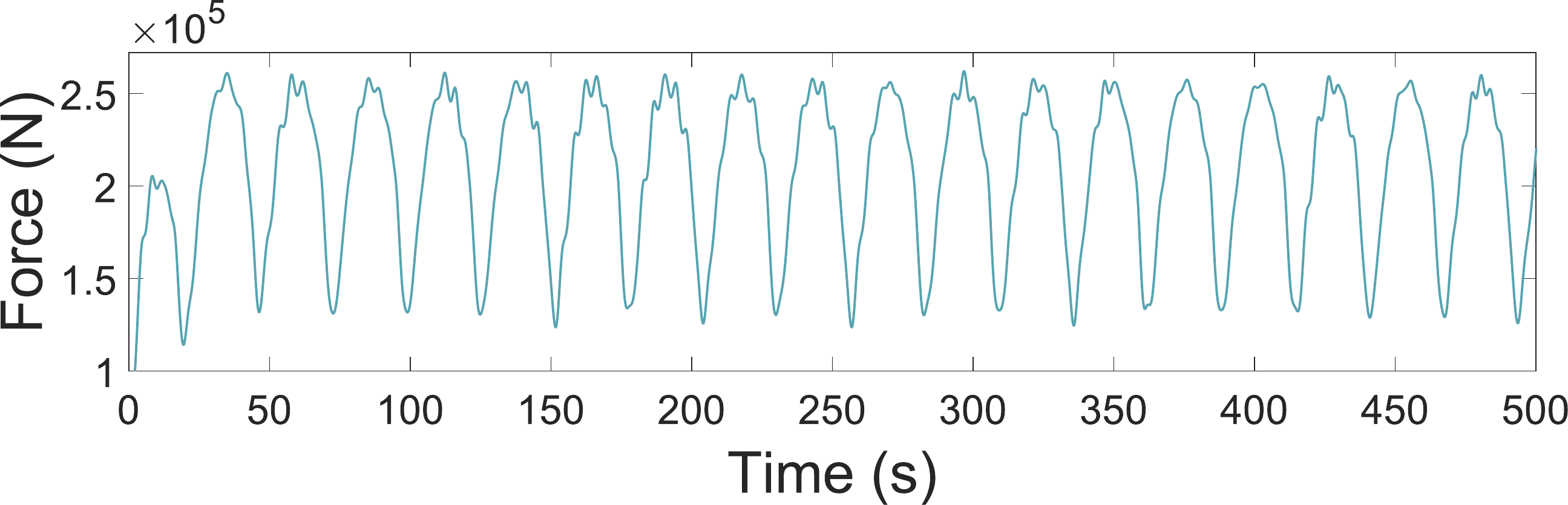}
        \caption{}
        \label{subfig:waveA_forceT}
    \end{subfigure}
    \\
    \begin{subfigure}{\textwidth}
    \centering
        \includegraphics[width=0.65\textwidth]{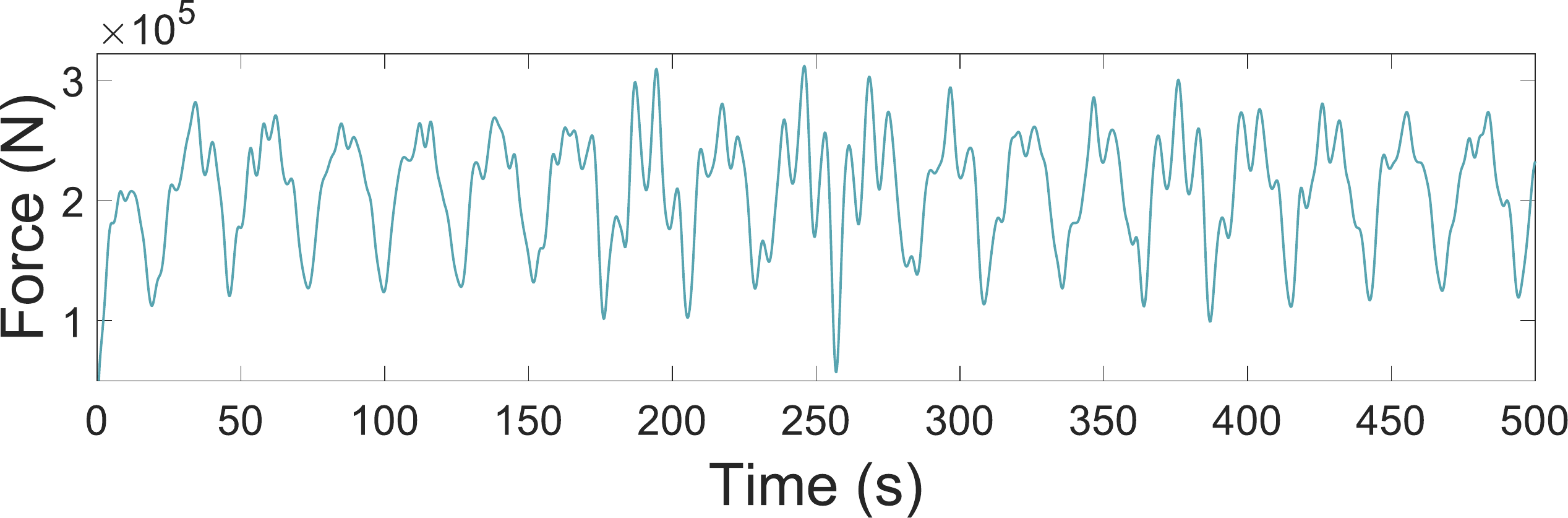}
        \caption{}
        \label{subfig:waveB_forceT}
    \end{subfigure}
    \caption{Pulling force in N with tether length $L=900\,$m.  (a) Onshore. (b) Offshore - Wave A. (c) Offshore - Wave B.}
    \label{subfig:total_forceT}
\end{figure}

\begin{table}[h]
\centering
\caption{$\eta$ values for increasing tether length in wave A scenario. $f_{traj}$ is the frequency of the kite's figure-eight paths under Algorithm \ref{alg:target point}}
\label{tab:eta}
\begin{tabular}{|c|c|c|c|c|}
\hline
$L$ (m) & $f_{traj}$ (Hz) & $F^t_{y_{peak}}$ (kN) & $y_{P_{peak}}$ (m)& \textbf{$\eta$} \\ \hline\hline
600              & 0.0324            & 100                   & 2                         & 0.02            \\ 
700              & 0.0287            & 93.5                  & 2.58                      & 0.0276          \\ 
800              & 0.0255            & 86                    & 3.27                      & 0.038           \\ 
900              & 0.023             & 78                    & 4                         & 0.0513          \\ 
1000              & 0.0208            & 72                    & 4.54                      & 0.0631          \\ 
1100             & 0.019             & 67                    & 4.6                       & 0.0687          \\ 
1200             & 0.0175            & 64                    & 4.33                      & 0.0677          \\ 
1300             & 0.0161            & 60                    & 3.9                       & 0.065           \\ \hline
\end{tabular}
\end{table}

\section{PROPOSED CONTROL APPROACH}\label{ch:control}

The analysis of Section \ref{ch:analysis} highlighted how the kite's trajectory can affect the platform oscillations, causing resonance effects which  could lead to increased fatigue and damage. To avoid this problem, we propose to modify the flight control approach by adjusting the length of the flown figure-eight trajectories on the basis of the kite's velocity, in order to indirectly control the resulting frequency of the tether force oscillations. While this general idea is relevant for any trajectory planning or navigation algorithm for AWES, in the specific case of Algorithm \ref{alg:target point} it can be realized by suitably adjusting the target points' locations, $P_-$ and $P_+$, as described in the remainder. The first step is to determine a desired path frequency, $f_{traj}^*$, that is far from the platform resonance peaks. Then, considering the average kite's speed over one figure-eight, denoted with  $\Bar{v}_K$, the desired trajectory length can be computed as:
\begin{equation}\label{eq:traj_length_des}
    L_{ traj}^*=\frac{\Bar{v}_K}{f_{traj}^*}
\end{equation}
The average kite speed can be estimated from onboard measurements or, to pre-tune the approach off-line, via simulations or the simplified equations of crosswind flight, see e.g. \cite{FaMiPi12}. 
On the other hand, with the considered switching law with two-target points (Algorithm \ref{alg:target point}), a rather accurate estimate of the trajectory length is given by:
\begin{equation}\label{eq:traj_length}
    L_{traj}=2\,(\Delta\theta+\Delta\phi)\,L
\end{equation}
where $\Delta\theta$ is the difference between maximum and minimum $\theta$ values experienced during the flight, $\theta_{max}$ and $\theta_{min}$, and $\Delta\phi=\phi_{+}-\phi_{-}$, see Fig. \ref{fig:kitePath-wave1} for an example. In the figure, note that the value $\theta_{min}\approx\theta_-$ can be assumed for up-loop trajectories (i.e., where the kite climbs on the sides and descends in the middle of the figure-eight), like the ones considered here. 
We can then estimate $\Delta\theta$ by considering the kite's turning radius $R$ during the figure-eight patterns, typically equal to about three times the kite's wingspan, so that the distance between the highest and lowest points of each turn, denoted by $\Delta z$, is
    \begin{equation}\label{eq:distance_radius}
        \Delta z \approx 2\,R.
    \end{equation}
At the same time, $\Delta z$ can be also estimated in the fixed reference frame as:
    \begin{equation}\label{eq:distance}
        \Delta z = L\,(\sin(\theta_{max})-\sin(\theta_-)).
    \end{equation}
 Combining \eqref{eq:distance_radius} and \eqref{eq:distance}, we get:
    \begin{equation}
        \theta_{max}=\arcsin\left(\frac{2\,R}{L}+\sin(\theta_-)\right)
    \nonumber
    \end{equation}
Using this formula, we obtain
\begin{equation}\label{eq:delta_theta}
\Delta\theta=\arcsin\left(\frac{2\,R}{L}+\sin(\theta_-)\right)-\theta_{min}
\end{equation}
We can finally determine the desired value of $\Delta\phi$, denoted $\Delta\phi^*$, by combining \eqref{eq:traj_length_des}, \eqref{eq:traj_length}, and \eqref{eq:delta_theta}:
\begin{equation}\label{eq:delta_phi}
\Delta\phi^*(t)=\frac{\Bar{v}_K}{2f_{traj}^*\,L(t)}-\arcsin\left(\frac{2\,R}{L(t)}+\sin(\theta_-)\right)+\theta_{min}
\nonumber
\end{equation}
Note that, for a fixed trajectory frequency, $\Delta\phi^*(t)$ decreases as the tether length $L(t)$ increases, as expected from the considerations reported at the end of Section \ref{ch:analysis}.\\
Then, it is sufficient to choose the target points such that $\phi_{+}-\phi_{-}=\Delta\phi^*$ to obtain with good approximation the desired path length and frequency. For example, if the target points' positions are symmetric to the $x$-axis, we have:
\begin{equation}
    |\phi_{+}(t)|=|\phi_{-}(t)|=\frac{\Delta\phi^*(t)}{2}
\nonumber
\end{equation}
and the target points to be used are $P_{-}=(\theta_{min},\phi_{-}(t))$ and $P_{+}=(\theta_{min},\phi_{+}(t))$. 


\section{SIMULATION RESULTS}\label{ch:simulation}

\begin{figure}
\centering
    \begin{subfigure}{0.45\textwidth}
        \captionsetup{justification=centering}
        \includegraphics[width=\textwidth]{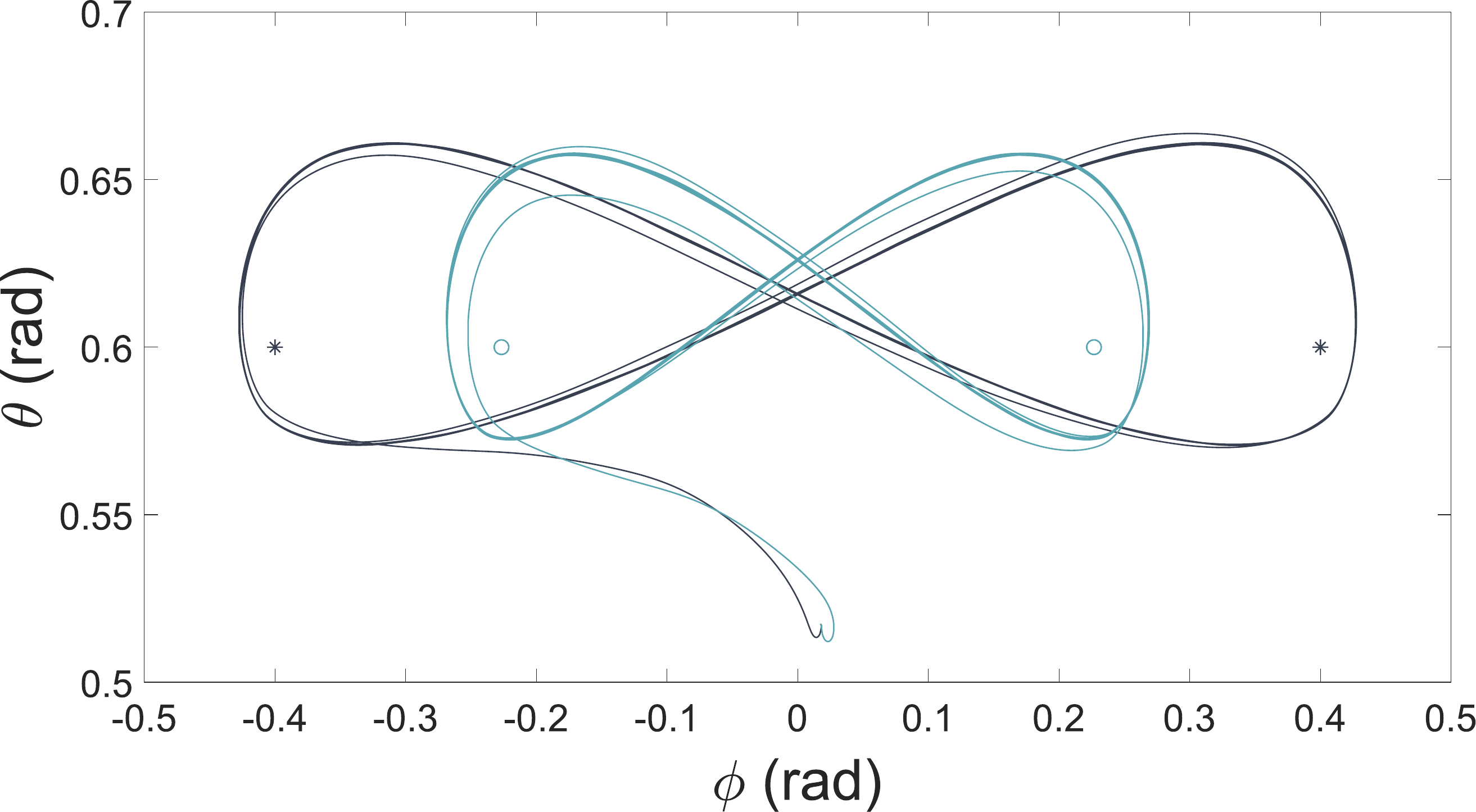}
        \caption{Kite's path in wave A scenario, with $L=1100\;m$. Kite trajectory and target points in the baseline approach (black lines and asterisks) and in the proposed approach to meet the desired oscillation frequency (light blue lines and circles).}
        \label{fig:kitePath-wave1}
    \end{subfigure}
\hspace{0.5cm}
    \begin{subfigure}{0.45\textwidth}
        \captionsetup{justification=centering}
        \includegraphics[width=\textwidth]{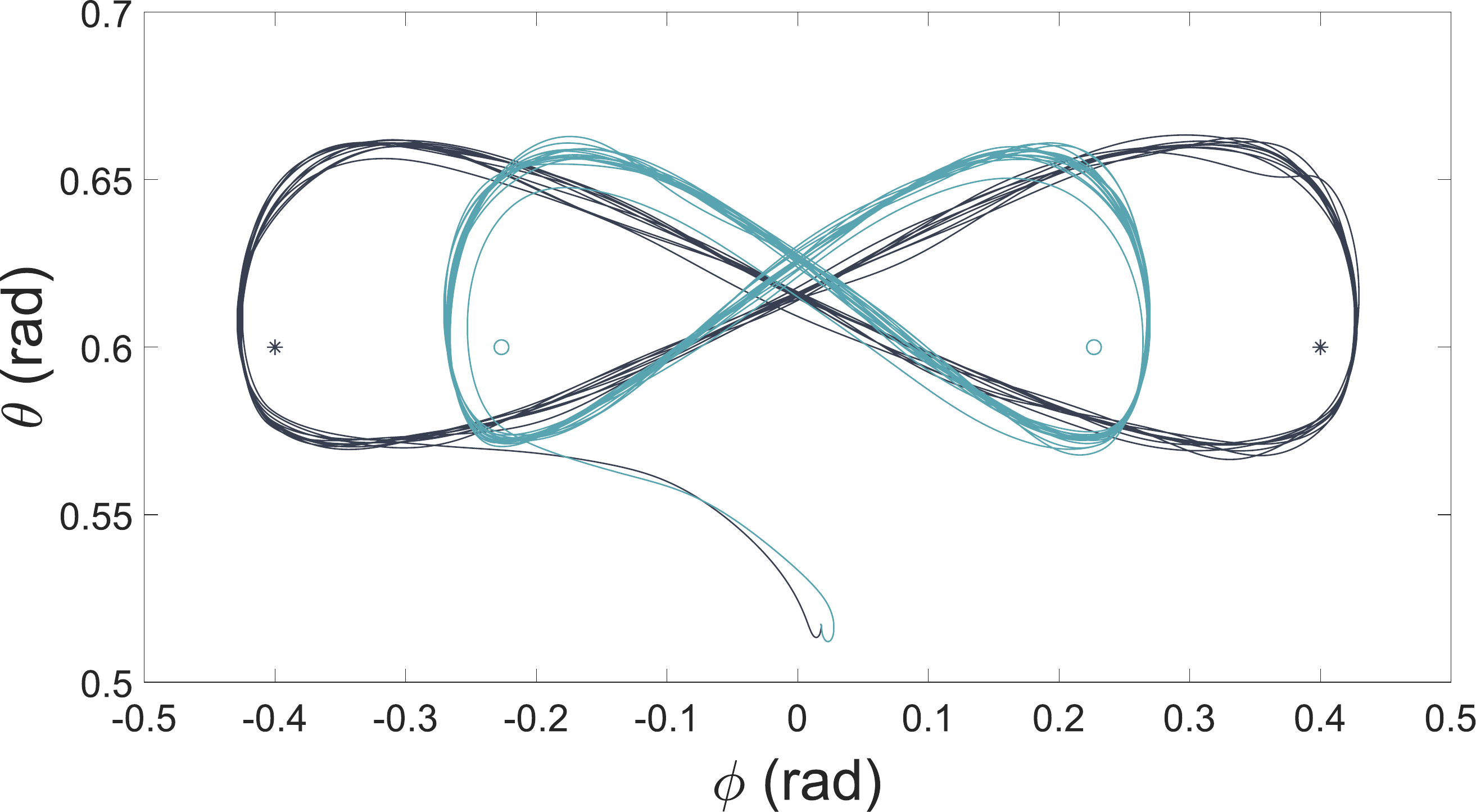}
        \caption{Kite's path in wave B scenario, with $L=1100\;m$. Kite trajectory and target points in the baseline approach (black lines and asterisks) and in the proposed approach to meet the desired oscillation frequency (light blue lines and circles).}
        \label{fig:kitePath-wave2}
    \end{subfigure}
\caption{Kite's path comparison.}
\end{figure}

We applied the proposed approach to control the frequency of force oscillations,  and compared the results to the baseline approach of fixed target points, considered in Section \ref{ch:analysis} (target points reported in Table \ref{tab:model}). In our approach, we chose a desired oscillation frequency of $f_{traj}^*=0.0305$~Hz, based on the analysis of Section \ref{ch:analysis}. Figs. \ref{fig:kitePath-wave1} and \ref{fig:kitePath-wave2} present a comparison between the kite patterns at two different tether lenghts and in wave scenarios A and B, respectively.  Besides noticing the different angular width of the trajectory in the $(\theta,\phi)$ plane, corresponding for the new approach to roughly the same linear width of 249~m  in the inertial plane, the figures also highlight the limited effects of waves on the kite trajectory in scenario A, while a slightly higher variability is found in scenario B, in line with the analysis of Section \ref{ch:analysis}.

\begin{figure}[b]
    \centering
    \includegraphics[width=0.65\textwidth]{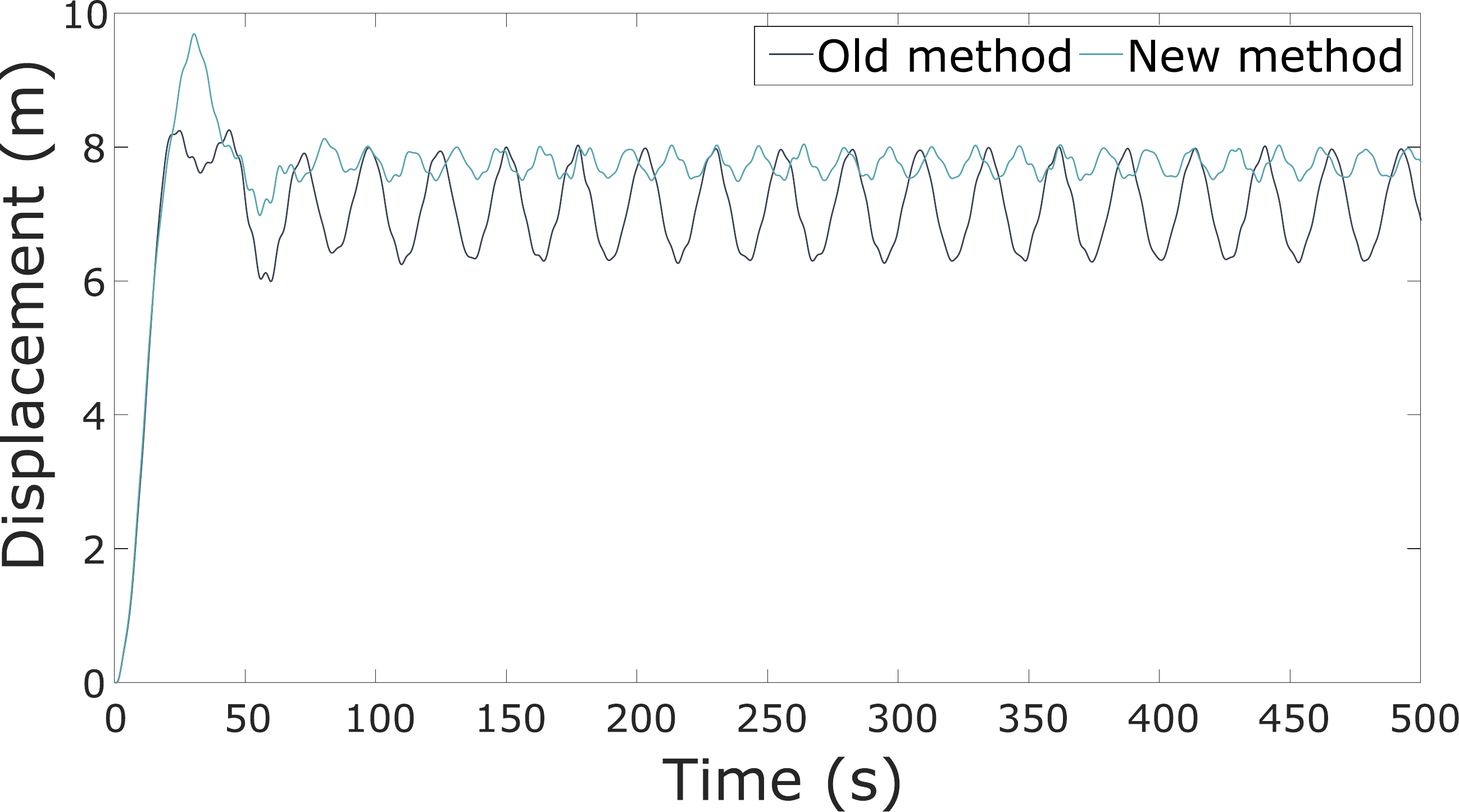}
    \caption{Platform surge motion in wave A scenario, using $L=1100\;m$.}
    \label{fig:SparMovementOld}
\end{figure}

\begin{figure}
    \centering
    \includegraphics[width=0.65\textwidth]{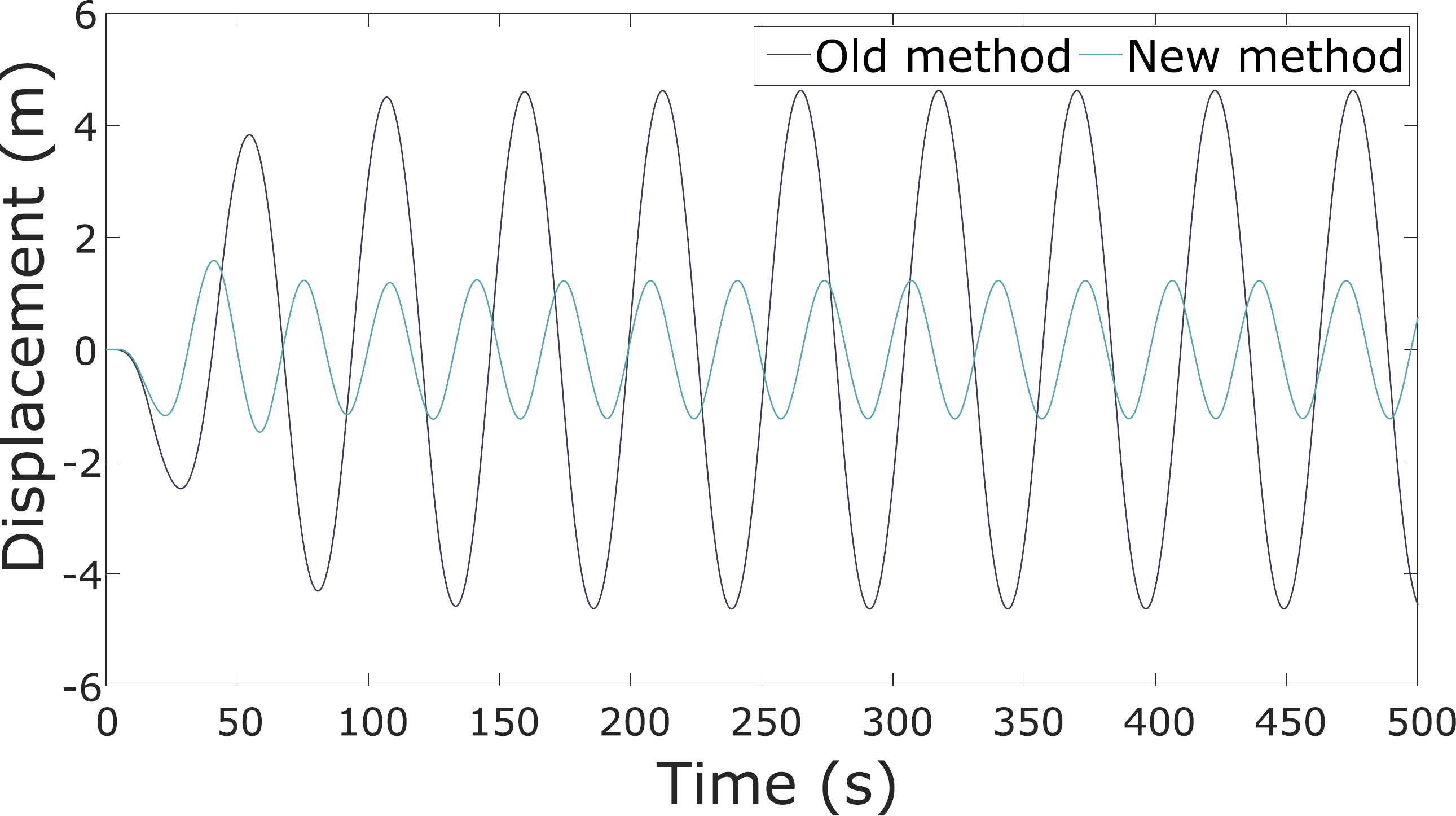}
    \caption{Platform sway motion in wave A scenario, using $L=1100\;m$.}
    \label{fig:SparMovementNew}
\end{figure}

\noindent Fig. \ref{fig:SparMovementOld} shows the spar's sway motion with $L=1100\;$m and wave A acting on the platform, with either fixed target points (baseline) or the proposed approach. Since the kite's frequency in this situation is approximately the same as the resonance (0.019 $Hz$ and 0.0185 $Hz$), the platform moves significantly in the baseline, while the amplitude of oscillations is reduced by more than half with our method.
In the surge direction, presented in Fig. \ref{fig:SparMovementNew}, after a first transient, a periodic motion at a frequency that is twice that of the sway oscillations. Such a  frequency corresponds to that of the $x$-component of traction force, which is twice that of the lateral component due to the figure-of-eight path shape. Also in this case, it can be noted that floater's oscillations are much more limited with the adjustment of target points depending on $L$.\\  
Regarding the heave direction, in both cases the frequency of oscillations is lower than the platform's resonance peak, so that the responses are similar and of small amplitude.\\
We finally recalculated $\eta$ \eqref{eq:eta} for the same tether length values and wave scenarios as those of Table \ref{tab:eta}, but with the new control strategy. The outcome, displayed in Table \ref{tab:eta_new}, demonstrates the method's effectiveness: the tether force oscillation frequency is kept far from the resonance, and $\eta$ remains roughly constant for the whole the reel-out phase. 

\begin{figure}
    \centering
    \includegraphics[width=0.65\textwidth]{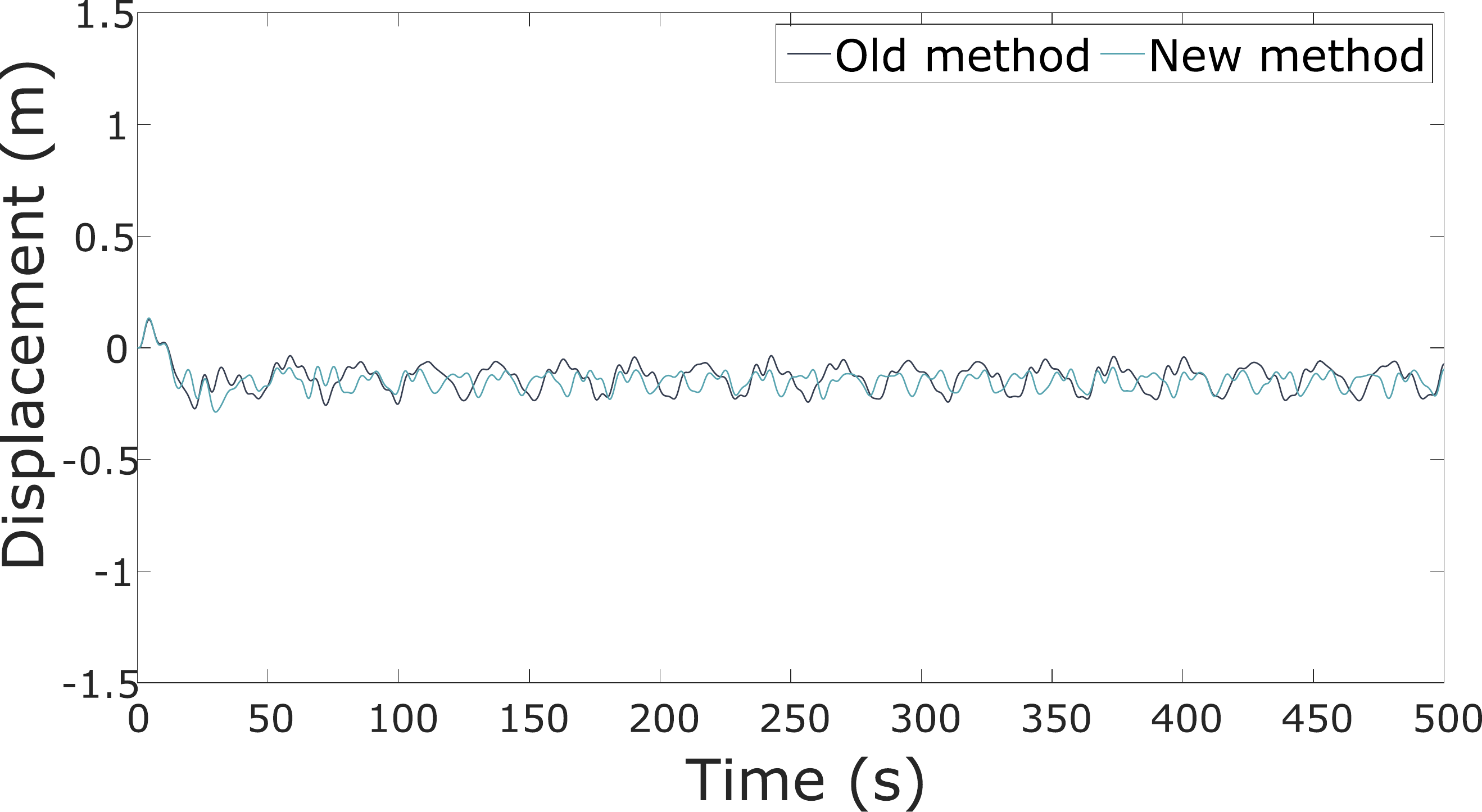}
    \caption{Platform heave motion in wave A scenario, using $L=1100\;m$.}
    \label{fig:SparMovementOldB}
\end{figure}

\begin{table}[H]
\centering
\caption{$\eta$ values with increasing tether length in wave A force scenario, using the proposed  control approach}
\label{tab:eta_new}
\begin{tabular}{|c|c|c|c|c|}
\hline
\textbf{L ($m$)} & \textbf{$f_{traj}$ ($Hz$)} & \textbf{$F^t_{y_{peak}}$ ($KN$)} & \textbf{$y_{P_{peak}}$ ($m$)} & \textbf{$\eta$} \\ \hline\hline
600              & 0.0324            & 100                   & 2                         & 0.02            \\ 
700              & 0.0318            & 88                    & 1.8                       & 0.0205          \\ 
800              & 0.0311            & 75.5                  & 1.67                      & 0.0221          \\ 
900              & 0.0308            & 65                    & 1.5                       & 0.0231          \\ 
1000             & 0.0305            & 58                    & 1.35                      & 0.0233          \\ 
1100             & 0.0302            & 51                    & 1.23                      & 0.0241          \\ 
1200             & 0.0298            & 45.5                  & 1.13                      & 0.0248          \\ 
1300             & 0.0295            & 40.5                  & 1.02                      & 0.0252          \\ \hline
\end{tabular}
\end{table}

\section{CONCLUSIONS AND FUTURE RESEARCH}\label{ch:conclusion}

The dynamical coupling between a deep offshore platform and an airborne wind energy system installed on it has been studied.
The analysis and simulation results demonstrate that the wing's path may be somewhat altered depending on the wave intensity; however, such a perturbation proved to be rather limited even with the highest waves tested.  Deepening the study for more wave types would undoubtedly be of interest. Future research may also concentrate on techniques for controlling the AWES winch to attenuate the effects of waves on the tether load.\\
On the other hand,  the analysis of the effects of the tether force on the platform motion highlighted a potential interference with the floater resonance peaks. 
A new control approach to keep the kite's trajectory frequency away from resonance has been presented and tested with promising results. The next steps of this research will be to investigate other platform configurations, to carry out wave tank tests for model validation,  and to study an integrated kite-platform design to limit the dynamic oscillations while optimizing relevant performance indicators, such as the amount of produced energy, its cost, and the environmental footprint of the technology. 

\bibliographystyle{IEEEtran}
\bibliography{bibliography}

\end{document}